\documentclass[12pt]{article}

\usepackage{amsmath}
\usepackage{feynmf}
\usepackage{latexsym}
\usepackage{amssymb}
\usepackage[dvips]{graphicx}
\usepackage{caption}
\usepackage{rotating}
\usepackage{latexsym}
\usepackage{verbatim}

 \def\be{\begin{equation}}
 \def\ee{\end{equation}}
 \def\bea{\begin{eqnarray}}
 \def\eea{\end{eqnarray}}
 \def\bei{\begin{itemize}}
 \def\eei{\end{itemize}}
 \def\bs{\begin{slide}}
 \def\es{\end{slide}}
 \def\nn{\nonumber}
 
 \def\pd{\partial}
 
 \def\L{\mathcal{L}}
 \def\M{\mathcal{M}}
 \def\a{\alpha}

 \def\g{\gamma}
 \def\G{\Gamma}
 \def\d{\delta}
 \def\D{\Delta}
 \def\m{\mu}
 \def\n{\nu}
 
 \def\r{\rho}
 \def\th{\theta}

 \def\l{\lambda}

 \def\s{\sigma}
 
 \def\sb{\bar{\s}}
 \def\psib{{\bar{\psi}}}
 \def\e{\epsilon}
 \def\f{\phi}

 \def\MZp{M_{Z'}}
 \def\MZO{M_{Z_0}}
 
 \def\({\left(}
 \def\){\right)}
 \def\[{\left[}
 \def\]{\right]}

 \def\Tr{\textnormal{Tr}}

 \def\thth{\theta^2 \bar\theta^2}

 \def\QHu{{Q_{H_u}}}
 \def\QHd{{Q_{H_d}}}
 \def\cA{\mathcal{A}}

 \newcommand{\bth}{{\bf 3}}
 \newcommand{\btw}{{\bf 2}}
 \newcommand{\bon}{{\bf 1}}

 \def\ds#1{#1\kern-1ex\hbox{/}}
\def\sla{\raise.15ex\hbox{$/$}\kern-.57em}

\begin{document}

\begin{titlepage}

\rightline{ROM2F/2009/09}

\vskip 2cm

\centerline{{\large\bf Gaugino radiative decay in an anomalous $U(1)'$ model}}

\vskip 1cm

\centerline{Andrea Lionetto\footnote{Andrea.Lionetto@roma2.infn.it}$^\natural$ and
            Antonio Racioppi\footnote{Antonio.Racioppi@kbfi.ee}$^\flat$}

\vskip 1cm

\centerline{$^\natural$ Dipartimento di Fisica dell'Universit\`a di Roma ,
``Tor Vergata" and}
\centerline{I.N.F.N.~ -~ Sezione di Roma ~ ``Tor Vergata''}
\centerline{Via della Ricerca  Scientifica, 1 - 00133 ~ Roma,~ ITALY}
\vskip 0.25cm
\centerline{$^\flat$ National Institute of Chemical Physics and Biophysics,}
\centerline{Ravala 10, Tallinn 10143, Estonia}
\vskip 1cm

%
%

\begin{abstract}
We study the neutralino radiative decay into the lightest
supersymmetric particle (LSP) in the framework of a minimal
anomalous
$U(1)'$ extension of the MSSM. It turns out that in a suitable decoupling
limit the
axino, which is present in the St\"uckelberg multiplet, is the LSP.
We compute the branching ratio (BR) for the decay of a neutralino into an axino and a photon.
We find that in a wide region of the parameter space, the BR is higher than 93\% in contrast with the typical value ($\lesssim 1\%$) in the CMSSM.
\end{abstract}

\end{titlepage}

\section{Introduction}

A great deal of work has been done recently to embed the standard model of
particle physics
(SM) into a brane construction
\cite{Marchesano:2007de,Blumenhagen:2005mu,Lust:2004ks,Kiritsis:2003mc}.
This research is part of the effort, initiated in \cite{Candelas:1985en},
to build a fully realistic four
dimensional vacuum out of string theory. While the original models were
formulated in the framework of the heterotic string,
the most recent efforts were formulated for type II strings in order to
take advantage of the recent work on
moduli stabilization using fluxes. Such brane constructions naturally lead
to extra anomalous $U(1)$'s in
the four dimensional low energy theory and, in turn, to the presence of
possible heavy $Z^\prime$ particles
in the spectrum. These particles should be among the early findings of LHC
and besides for the above cited
models they are also a prediction of many other theoretical models of the
unification of forces (see \cite{Langacker:2008yv} for a recent review).
It is then of some interest to know if
these $Z^\prime$ particles contribute to the cancellation of the gauge
anomaly in the way predicted from string theory or not.
In \cite{ourpaper} some of the present authors have studied a
supersymmetric (SUSY) extension of the minimal
supersymmetric standard model (MSSM) in which the anomaly is cancelled
{\it \`a la} Green-Schwarz.
The model is only string-inspired and is not the low-energy sector of some
brane construction. The reason of this choice
rests in our curiosity to explore the phenomenology of these models keeping
a high degree of flexibility,
while avoiding the intricacies and uncertainties connected with a string
theory construction.
For previous work along these lines we refer to
\cite{Anastasopoulos:2006cz,Coriano':2005js,Coriano:2006xh,Coriano:2007fw,Coriano:2007xg,Armillis:2008vp,Kors:2004ri,Kors:2005uz,Feldman:2006wd,DeRydt:2007vg}.
In this work we continue the analysis of the axino interactions \cite{Fucito:2008ai},
studying the neutralino radiative decay into the axino.
The next to lightest supersymmetric particle (NLSP) radiative decay
might be the first process where the LSP could be observed at LHC
\cite{Baer:2002kv,Dreiner:2006sb,Basu:2007ys,Baer:2008ih}.
In our model we assume the axino as the LSP and a generic MSSM neutralino as the NLSP.
Due to the axino interaction vertices the NLSP neutralino can only decay into
an axino plus either a photon or a $Z_0$ or SM fermions.
We compute all the amplitudes for that decays and the BR for the decay of a neutralino into an axino and a photon.
This is the plan of the paper:
in Section 2 we describe our model. In Section 3 and 4 we find the LSP and
study the axino interactions. In Section 5 we compute all the decay rates under study and finally in Section 6
we show the results and the related plots.
Section 7 is devoted to the conclusions.

\section{Model Setup} \label{sect:Model Setup}

In this section we briefly discuss our theoretical framework. We assume an
extension
of the MSSM with an additional abelian vector multiplet $V^{(0)}$ with
arbitrary
charges. The anomalies are cancelled with the Green-Schwarz (GS) mechanism
and with
the Generalized Chern-Simons (GCS) terms. All the details can be found
in~\cite{ourpaper}.
All the MSSM fields are charged under the additional vector multiplet
$V^{(0)}$,
with charges which are
given in Table~\ref{QTable}, where $Q_i, L_i$ are the left handed quarks
and leptons
respectively while $U^c_i,
D^c_i, E^c_i$ are the right handed up and down quarks and the electrically
charged
leptons. The superscript $c$ stands
for charge conjugation. The index $i=1,2,3$ denotes the three different
families.
$H_{u,d}$ are the two Higgs
scalars.
  \begin{table}[h]
  \centering
  \begin{tabular}[h]{|c|c|c|c|c|}
   \hline & SU(3)$_c$ & SU(2)$_L$  & U(1)$_Y$ & ~U(1)$^{\prime}~$\\
   \hline $Q_i$   & $\bth$       &  $\btw$       &  $1/6$   & $Q_{Q}$ \\
   \hline $U^c_i$   & $\bar \bth$  &  $\bon$       &  $-2/3$  & $Q_{U^c}$
\\
   \hline $D^c_i$   & $\bar \bth$  &  $\bon$       &  $1/3$   & $Q_{D^c}$
\\
   \hline $L_i$   & $\bon$       &  $\btw$       &  $-1/2$  & $Q_{L}$ \\
   \hline $E^c_i$   & $\bon$       &  $\bon$       &  $1$     &
$Q_{E^c}$\\
   \hline $H_u$ & $\bon$       &  $\btw$       &  $1/2$   & $Q_{H_u}$\\
   \hline $H_d$ & $\bon$       &  $\btw$       &  $-1/2$  & $Q_{H_d}$ \\
   \hline
  \end{tabular}
  \caption{Charge assignment.}\label{QTable}
  \end{table}

The key feature of this model is the mechanism of anomaly cancellation.
As it is well known, the MSSM is anomaly free. In our MSSM extension all
the anomalies
that involve only the $SU(3)$, $SU(2)$ and $U(1)_Y$ factors vanish
identically. However, triangles with $U(1)'$ in the external legs
in general are potentially anomalous.
These anomalies are\footnote{We are working in an effective field theory
framework
and we ignore throughout the paper all the gravitational effects.  In
particular, we
do not consider the gravitational anomalies which, however, could be
canceled by the
Green-Schwarz mechanism. Moreover it can be shown that $\cA^{(3)}=0$ \cite{ourpaper}. }
\bea
   U(1)'-U(1)'-U(1)':     &&\ \cA^{(0)}  \\
   U(1)'-U(1)_Y - U(1)_Y: &&\ \cA^{(1)} \\
   U(1)'-SU(2)-SU(2):     &&\ \cA^{(2)} \\
   U(1)'-SU(3)-SU(3):     &&\ \cA^{(3)} \\
   U(1)'-U(1)'-U(1)_Y:    &&\ \cA^{(4)}
\eea
All
the remaining anomalies that involve $U(1)'$s
vanish identically due to group theoretical arguments
(see Chapter 22 of \cite{Weinberg2}).
Consistency of the model is achieved by the contribution of a
St\"uckelberg field
$S$ and its appropriate couplings to the anomalous $U(1)'$. The
St\"uckelberg lagrangian written in terms of superfields is
\cite{Klein:1999im}
  \be
 \L_S = {\frac{1}{4}} \left. \( S + S^\dagger + 4 b_3 V^{(0)} \)^2
\right|_{\thth}\!\!
                - {\frac{1}{2}} \left\{ \[\sum_{a=0}^2 b^{(a)}_2 S
\Tr\( W^{(a)}
W^{(a)} \) + b^{(4)}_2 S W^{(1)} W^{(0)} \]_{\th^2} \!\!\!\!\!\!+h.c.\!
\right\}
   \label{Laxion}
\ee
where the index $a=0,\ldots,2$ runs over the $U(1)',\, U(1)_Y$ and $SU(2)$ gauge groups respectively.
The St\"uckelberg multiplet is a chiral superfield
 \be
    S =  s+ i\sqrt2 \th \psi_S + \th^2 F_S - i \th \s^\m \bar\th \pd_\m s
+
                {\frac{\sqrt2}{2}}  \th^2 \bar\th \bar\s^\m \pd_\m \psi_S
-
{\frac{1}{4}} \thth \Box s \label{Smult}
 \ee
and transforms under $U(1)'$ as
\bea
   V^{(0)} &\to& V^{(0)} + i \( \Lambda - \Lambda^\dag \) \nn\\
   S  &\to& S - 4 i ~b_3 ~\Lambda \label{U1'}
  \label{U1Trans}\eea
where $b_3$ is a constant related to the $Z'$ mass.
In our model there are two mechanisms that give mass to the gauge bosons:
(i) the
St\"uckelberg mechanism and (ii) the Higgs mechanism.
In the following we assume\footnote{We impose this condition to simplify our computations
and to give a compact analytical expressions. There are no obstructions to
set $\QHu\neq 0, \, \QHd\neq 0$.} that
\be
 \QHu=\QHd=0 \label{QHu}
\ee
The mass terms for the gauge fields are given by
     \be
      \L_M = \frac{1}{2} \(V^{(0)}_\m \ V^{(1)}_\m \ V^{(2)}_{3\m} \) M^2
                          \( \begin{array}{c} V^{(0)\m}\\ V^{(1)\m}\\
V^{(2)\m}_{3}
\end{array} \)
     \ee
with $M^2$ being the gauge boson mass matrix
     \be
      M^2= \( \begin{array}{ccc} M_{V^{(0)}} & ~~~0& ~~~0  \\
                 ... & g_1^2 \frac{v^2}{4} & -g_1 g_2 \frac{v^2}{4}   \\
                 ... & ...  & g_2^2 \frac{v^2}{4}  \\\end{array} \)
  \label{BosonMasses} \ee
where $M_{V^{(0)}}=4 b_3 g_0$  is the mass parameter for the anomalous
$U(1)$ and it
is assumed to be in the TeV
range. The lower dots denote the obvious terms under symmetrization.
After diagonalization, we obtain the eigenstates
 \bea
   A_\m &=&\frac{g_2 V^{(1)}_\m + g_1 V^{(2)}_{3\m}}{\sqrt{g_1^2+g_2^2}}
\label{photon}\\
    Z_{0\m} &=& \frac{g_2 V^{(2)}_{3\m} - g_1
V^{(1)}_\m}{\sqrt{g_1^2+g_2^2}}
\label{Z0}\\
   Z'_\m  &=& V^{(0)}_\m \label{Zprime}
    \eea
and the corresponding masses
   \bea
   M^2_{\g}&=&0\\
    M^2_{Z_0} &=&\frac{1}{4} \(g_1^2+g_2^2\) v^2\label{Z0mass}\\
M^2_{Z'}  &=&M_{V^{(0)}}^2
  \label{Zpmass} \eea
Finally the rotation matrix from the hypercharge to the photon
basis is
    \bea
     \( \begin{array}{c} Z'_\m\\
                         Z_{0 \m}\\
                         A_\m \end{array} \)
       &=&O_{ij}
     \( \begin{array}{c} V^{(0)}_\m\\
                         V^{(1)}_\m\\
                         V^{(2)}_{3\m} \end{array} \)\label{Oij}\\
      &=&\( \begin{array}{ccc}      1& 0  ~~& 0  \\
                                0& - \frac{g_1}{{\sqrt{g_1^2+g_2^2}}}
& \frac{g_2}{{\sqrt{g_1^2+g_2^2}}}  \\
                                0 & \frac{g_2}{{\sqrt{g_1^2+g_2^2}}} &
\frac{g_1}{{\sqrt{g_1^2+g_2^2}}}  \\ \end{array} \)
     \( \begin{array}{c} V^{(0)}_\m\\
                         V^{(1)}_\m\\
                         V^{(2)}_{3\m} \end{array} \) \nn \eea
where $i,j=0,1,2$.

We now give the expansion of the lagrangian piece $\L_S$ defined in (\ref{Laxion}) in component fields only for the part that
is needed in the following sections. Using the Wess-Zumino gauge we get
\bea
 \L_{axino}&=&{\frac{i}{4}} \psi_S \s^\m \pd_\m \psib_S -\sqrt2 b_3 \psi_S
\l^{(0)}
   -{\frac{i}{2\sqrt2}} \sum_{a=0}^2b^{(a)}_2  \Tr \( \l^{(a)} \s^\m
\sb^\n F_{\m
\n}^{(a)} \) \psi_S \nn\\
   &&-{\frac{i}{2\sqrt2}} b^{(4)}_2 \[ {\frac{1}{2}} \l^{(1)} \s^\m \sb^\n
F_{\m
\n}^{(0)} \psi_S
   + (0 \leftrightarrow 1) \] +h.c.
\label{axinolagr}\eea
As it was pointed out in \cite{Anastasopoulos:2006cz}, the
St\"uckelberg mechanism is not enough to cancel all the
anomalies. Mixed anomalies between anomalous and non-anomalous
factors require an additional mechanism to ensure consistency of
the model: non gauge invariant GCS terms must be added.
In our case, the GCS terms have the form
\cite{Andrianopoli:2004sv}
   \bea
 \L_{GCS} &=&- d_4     \[ \( V^{(1)} D^\a V^{(0)} - V^{(0)} D^\a V^{(1)}\) W^{(0)}_\a + h.c. \]_{\thth} +\nn\\
        &&+  d_5     \[ \( V^{(1)} D^\a V^{(0)} - V^{(0)} D^\a V^{(1)}\) W^{(1)}_\a + h.c. \]_{\thth} +\nn\\
       &&+  d_6 \Tr \bigg[ \( V^{(2)} D^\a V^{(0)} - V^{(0)} D^\a V^{(2)}\) W^{(2)}_\a + n.a.c + h.c. \bigg]_{\thth}
   \label{GCS_1}
\eea
where $n.a.c.$ refers to non abelian completion terms.
The $b$ constants in (\ref{Laxion}) and the $d$ constants in (\ref{GCS_1}) are fixed by the anomaly cancellation procedure
(for details see~\cite{ourpaper}).

For a symmetric distribution of the anomaly, we have
\bea
 &&     b^{(0)}_2 b_3 =-\frac{\cA^{(0)}}{384\pi^2}~
 \qquad b^{(1)}_2 b_3 = - \frac{\cA^{(1)}}{128 \pi^2}
 \qquad b^{(2)}_2 b_3 = -\frac{\cA^{(2)}}{64 \pi^2}~
 \qquad b^{(4)}_2 b_3 = -\frac{\cA^{(4)}}{128 \pi^2}\nn\\
 &&~~~~ d_4 =- \frac{\cA^{(4)}}{384 \pi^2}~~~~~
 \qquad d_5 = \frac{\cA^{(1)}}{192\pi^2}~~~~~~
 \qquad d_6= \frac{\cA^{(2)}}{96 \pi^2}
\label{bsds}  \eea
It is worth noting that the GCS coefficients $d_{4,5,6}$ are fully
determined in terms of the $\cA$'s by the gauge invariance, while the
$b_2^{(a)}$'s
depend
only on the free parameter $b_3$, which is related to the mass of
the anomalous $U(1)$.

The soft breaking sector of the model is given by
   \be
    \L_{soft}=\L_{soft}^{MSSM}- {\frac{1}{2}}  \(M_0 \l^{(0)} \l^{(0)} + h.c. \)
    - {\frac{1}{2}}  \(\frac{M_S}{2} \psi_S \psi_S  + h.c. \) \label{Lsoft}
   \ee
where $\L_{soft}^{MSSM}$ is the usual soft susy breaking lagrangian while
 $\l^{(0)}$ is the gaugino of the added
$U(1)'$ and $\psi_S$ is the axino.
The axino soft mass term deserves some comment:
from~\cite{Girardello:1981wz} we know that a fermionic mass term for a chiral multiplet is not allowed in presence
of Yukawa interactions in which this chiral multiplet is involved.
But in the classical Lagrangian the St\"uckelberg multiplet cannot contribute to superpotential terms
given that the gauge invariance
given from our U(1)' symmetry (\ref{U1Trans}) requires non-holomorphicity in the chiral fields.
In fact in our model both the axino and the axion couple only through GS interactions.
It is worth noting that a mass term for the axion $\f$ is instead not allowed since it
transforms non trivially under the anomalous $U(1)'$ gauge transformation~(\ref{U1Trans}).
 At first sight our lagrangian may look not the most general possible one.
In particular, an explicit Fayet-Iliopoulos (FI) term $\xi V^{(0)}$ and an explicit kinetic mixing term $\d W^{(0)} W^{(1)}$
could be added.
For what concerns the FI term, it is well known that in certain string-inspired models (see, e.g.
\cite{Poppitz:1998dj,Kiritsis:2008ry}), a one loop FI term is absent, even if
$\Tr(Q) \neq 0$. This  is in  apparent conflict with the observation
\cite{Fischler:1981zk} that in field theory a quadratically divergent FI term is
always generated at one loop.
The solution to this paradox is that in the low-energy lagrangian there
should be a counterterm, which compensates precisely, i.e. both the
divergent and the finite part of the one loop contribution.
We do not write explicitly this counterterm, since its exact expression is
model and regularization dependent, but we implicitly assume that such a
cancellation occurs.
For what concerns the kinetic mixing term, also this arises at one loop level and $\d \propto \Tr(Q Y)$.
 To simplify our computations we cancel the $\d$ term by choosing
\be
 Q_Q = Q_L \label{QQequalQL}
\ee
Moreover the constraint (\ref{QHu}) implies $\cA^{(4)}=0$, cancelling another possible source of kinetic mixing.
Finally, as mentioned before, we do not take into account the anomaly cancellation in the gravitational sector.

\section{Neutralino Sector \label{Neutralinos}}
Assuming the conservation of R-parity the LSP is a good weak interacting massive particle (WIMP) dark matter candidate.
As in the MSSM the LSP is given by a linear combination of fields in the
neutralino sector.
The general form of the neutralino mass matrix is given
in~\cite{ourpaper}. Written
in the interaction eigenstate basis
$(\psi^{0})^T= (\psi_S, \ \l^{(0)},\ \l^{(1)} ,\ \l^{(2)}_3,\ \tilde h_d^0,\ \tilde
h_u^0)$
it is a six-by-six matrix.
From the point of view of the strength of the interactions the two extra
 states are
not on the same footing with respect
to the standard ones. The axino and the extra gaugino $\l^{(0)}$ dubbed primeino are in fact extremely weak interacting massive particle (XWIMP).
Thus we are interested in situations in which the extremely weak sector is
decoupled
from the standard one and the LSP
belongs to this sector.
This can be achieved at tree level with the choice (\ref{QHu}).
The neutralino mass matrix $ {\bf M}_{\tilde N} $ becomes
   \be
    {\bf M}_{\tilde N}
     =   \(\begin{array}{cccccc}
          \frac{M_S}{2} & \frac{M_{V^{(0)}}}{\sqrt2} &   0   &   0   &
     0
    & 0 \\
                  \dots &       M_0                  &   0   &   0   &
     0
    & 0   \\
                  \dots &      \dots                 &  M_1  &   0   &
-\frac{g_1
v_d}{2}& \frac{g_1 v_u}{2} \\
                  \dots &      \dots                 & \dots &  M_2  &
\frac{g_2
v_d}{2} & -\frac{g_2 v_u}{2} \\
                  \dots &      \dots                 & \dots & \dots &
     0
    & -\m  \\
                  \dots &      \dots                 & \dots & \dots &
   \dots
    & 0
         \end{array}\) \label{massmatrix} ~~~~
   \ee
where $M_S,~M_0,~M_1,~M_2$ are the soft masses coming from the soft
breaking terms
(\ref{Lsoft}) while $M_{V^{(0)}}$ is given in
~(\ref{BosonMasses}).
It is worth noting that the D terms and kinetic mixing terms
can be neglected in the tree-level computations of the eigenvalues and
eigenstates.

Moreover we make the assumption that $M_0 \gg M_S,M_{V^{(0)}}$ so the eigenstates are nearly pure axino and primeino,
and we suppose that the axino $\psi_S$ is the LSP \cite{Fucito:2008ai}. We consider the NLSP to be either
a mixture of the bino $\l^{(1)}$ and higgsino or a nearly pure wino~$\l^{(2)}_3$.
The first situation is a typical configuration of the mSUGRA parameter space\footnote{Or in the so called Constrained MSSM (CMSSM)}
while the second situation is naturally realized in anomaly mediated supersymmetry breaking scenarios.

\section{Axino Interactions}
\begin{figure}[t]
\centering
\includegraphics[scale=0.35]{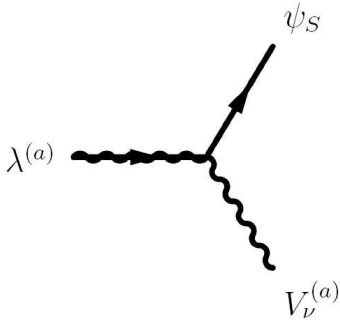}
      \caption{Gaugino-axino-vector interaction vertex.}
\label{laxiY}
\end{figure}

 The axino interactions can be read off from the interaction
lagrangian~(\ref{axinolagr}). The relevant term, written in terms of four
components Majorana spinors\footnote{The gamma matrices $\gamma^\mu$ are
in the
Weyl representation.}, is given by
\be
 \L =i \sqrt{2} g_1^2 b_2^{(1)}  \bar\Lambda^{(1)} \g_5 [\g^\m,\g^\n](\partial_\m V^{(1)}_\n)    \Psi_S +
        i {\sqrt{2}\over2} g_2^2 b_2^{(2)}  \bar\Lambda^{(2)}_3 \g_5 [\g^\m,\g^\n](\partial_\m V^{(2)}_{3\n}) \Psi_S
\ee
where the $b_2^{(a)}$ coefficients are given in~(\ref{bsds}). The related interaction vertex is depicted in Fig.~\ref{laxiY}
and the corresponding Feynman rule is
\be
 \ C^{(a)} [\g^\m,\g^\n]ik_\m \label{lpg}
 \ee
where $k_\m$ is the momentum of the outgoing vector and $C^{(a)}$'s
\bea
 C^{(1)}&=&\sqrt{2}g_1^2 b_2^{(1)} \nn\\
 C^{(2)}&=&{\sqrt{2}\over2} g_2^2 b_2^{(2)}
 \label{vertice}
\eea
are factors which contain the coupling constants and the parameters $b_2^{(a)}$ which are
related to the anomalous $U(1)'$ \cite{ourpaper}. Therefore $C^{(a)} \ll g_a$ and
the axino interactions will be extremely weak, being suppressed by an order of
magnitude factor with respect to the weak interactions.

\section{Gaugino Decay Channels}

\begin{figure}[t]
\centering
\includegraphics[scale=0.4]{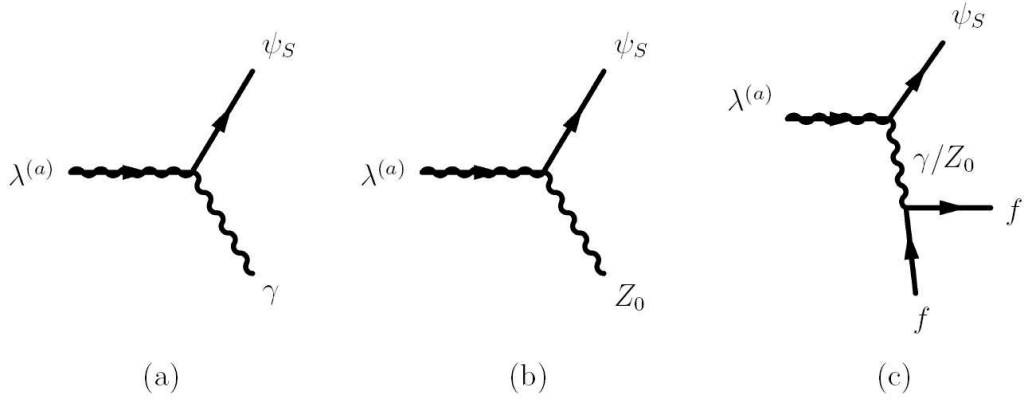}
      \caption{Gaugino decay modes. (a) Axino-photon production (b) Axino-$Z_0$ production (c)~Axino-$f \bar f$ pair production.}
\label{binodecay}
\end{figure}

In our model there are only few allowed decay channels.
By assuming R parity conservation, the NLSP can decay only into the LSP plus other SM particles.
Since the axino (LSP) interacts only with vertices of Fig.~\ref{laxiY},
only the gaugino fraction of the neutralino (NLSP) gives a contribution. It can decay at the leading order
only into gauge vector bosons or SM fermions (see Fig.~\ref{binodecay}). Moreover, since we assume a near
mass degeneracy between $\psi_S$ and the NLSP, the production of the $t \bar t$ pair is suppressed,
while the production of the $Z_0$ is allowed only for high neutralino masses.
To emphasize the fact that only the gaugino component plays a key role in the radiative decay, from now on
we will refer to gaugino instead of neutralino. For the cases in hand (bino-higgsino or pure wino NLSP)
the BR of the process will be independent on the gaugino
fraction since this is factorized in all the amplitude.

\subsection{$\l^{(a)} \to \psi_S \g$ decay rate}
The corresponding Feynman diagram is depicted in Fig.~\ref{binodecay}a. The decay rate is given by
\be
 \G_\g^{(a)} = \frac{k_\g}{32\pi^2 M_a^2} \int |\M_\g^{(a)}|^2 \, d\Omega
\ee
where
\be
 k_\g=\frac{M_a^2-M_S^2}{2 M_a}
\ee
is the purely spatial momentum of the outgoing photon in the gaugino rest frame, $M_a$ is the gaugino mass and $M_S$ is the axino
mass and
\be
 \M_\n^{(a)} =i  C_\g^{(a)} \bar u_S \g_5 \[ \g_\m,\g_\n \] u_\l k^\m
\ee
is the Feynman amplitude of the process with $k^\m$ the outgoing photon momentum.
The following coefficients
\bea
 C^{(1)}_\g&=&\sqrt2 g_1^2 b_2^{(1)} \cos\theta_W \nn\\
 C^{(2)}_\g&=&{\sqrt2\over2} g_2^2 b_2^{(2)} \sin\theta_W \label{Cg}
\eea
are the factors $C^{(a)}$ rotated on the photon eigenstate.
The square amplitude is given by
\bea
 |\M_\g^{(a)}|^2 &=& \frac{1}{2} \sum_{\l} \e^{\s}_{(\l)} \e^{* \n}_{(\l)} \ \M_{\s}^{(a)*} \M_{\n}^{(a)}
                 = - \frac{1}{2} \eta^{\s \n} \ \M_{\s}^{(a)*} \M_{\n}^{(a)} \nn\\
           &=& - \frac{\(C_\g^{(a)}\)^2}{2} \eta^{\s \n} k^\r k^\m T_{\m\n\s\r}
\eea
where
\be
 T_{\m\n\s\r}=\Tr \big[ \ds p_S \[\g_\m,\g_\n\] \ds p_a \[\g_\s,\g_\r\] \big]- M_S M_a \Tr \big[  \[\g_\m,\g_\n\] \[\g_\s,\g_\r\] \big]
 \label{Trace}
\ee
with $p_S$ and $p_a$ the axino and gaugino momenta respectively.
The pre-factor $1/2$ is the average over the spin states of the gaugino. Performing the computations we finally get
\be
 \G_\g^{(a)} = \(g_0 \cA^{(a)}\)^2 g_a^4 R^{(a)}_\g \frac{(\D M)^3 (\D M+2 M_S)^3}{1024 \pi^5 (\D M +M_S)^3 \MZp^2}
\ee
where $\D M=M_a-M_S$, $\MZp$ the $Z'$ mass, $\cA^{(a)}$ is the anomaly factor and
\bea
 R^{(1)}_\g=\cos^2\theta_W \nn\\
 R^{(2)}_\g=\sin^2\theta_W \label{Rg}
\eea
\subsection{$\l^{(a)} \to \psi_S Z_0$ decay rate}
The corresponding Feynman diagram is depicted in Fig.~\ref{binodecay}b. The situation is very similar to the previous case except that we have a massive vector in the final state.
The decay rate is given by
\be
 \G_{Z_0}^{(a)} = \frac{k_{Z_0}}{32\pi^2 M_a^2} \int |\M_{Z_0}^{(a)}|^2 \, d\Omega
\ee
where
\be
 k_{Z_0}=\sqrt{\( \frac{M_a^2-M_S^2+\MZO^2}{2 M_a} \)^2 - \MZO^2}
\ee
is the purely spatial momentum of the outgoing $Z_0$ in the gaugino rest frame.
The square modulus of the amplitude is given by
\bea
 |\M_{Z_0}^{(a)}|^2 &=& \frac{1}{2} \sum_{\l} \e^{\s}_{(\l)} \e^{* \n}_{(\l)} \ \M_{\s}^{(a)*} \M_{\n}^{(a)}
           = \frac{1}{2} \( - \eta^{\s \n} + \frac{k^\s k^\n}{\MZO^2}\) \M_{\s}^{(a)*} \M_{\n}^{(a)} \nn\\
           &=& \frac{\(C_{Z_0}^{(a)}\)^2}{2}  \( - \eta^{\s \n} + \frac{k^\s k^\n}{\MZO^2}\) k^\r k^\m T_{\m\n\s\r}
\eea
where
\bea
 C_{Z_0}^{(1)}&=&-\sqrt2 g_1^2 b_2^{(1)} \sin\theta_W  \nn\\
 C_{Z_0}^{(2)}&=&{\sqrt2\over2} g_2^2 b_2^{(2)} \cos\theta_W  \label{CZ0}
\eea
are the factor $C^{(a)}$ rotated on the $Z_0$ eigenstate while $T_{\m\n\s\r}$ is given in (\ref{Trace}).
The pre-factor $1/2$ is related to the average over the spin states of the gaugino.
We finally get
\bea
 \G_{Z_0}^{(a)} &=& \( g_0 A^{(a)} \)^2 g_a^4 R^{(a)}_{Z_0}
            \left\{\frac{ \[ \(\D M\)^2-\MZO^2 \] \((\D M+2 M_S)^2-\MZO^2\)}{(\D M+M_S)^2}\right\}^{3/2} \times \nn\\
            &&\times \frac{ \(2 \(\D M\)^2+\MZO^2\)}{2048 \pi^5 (\[ \(\D M\)^2-\MZO^2 \] \MZp^2}
\eea
where $\D M=M_a-M_S$, $\MZp$ the $Z'$ mass, $\cA^{(a)}$ is the anomaly factor and
\bea
 R^{(1)}_{Z_0}=\sin^2\theta_W \nn\\
 R^{(2)}_{Z_0}=\cos^2\theta_W \label{RZ}
\eea
\subsection{$\l^{(a)} \to \psi_S f \bar f$ decay rate}
The decay is kinematically allowed only if $\D M > 2 m_f$. The corresponding Feynman diagram is depicted in Fig.~\ref{binodecay}c.
The decay rate is given by
\be
 \G_{f \bar f}^{(a)} = \frac{1}{\( 2 \pi\)^3} \frac{1}{32 M_a^3}
                 \int^{m_{12}^\text{2 max}}_{m_{12}^\text{2 min}} d m_{12}^2
                 \int^{m_{23}^\text{2 max}}_{m_{23}^\text{2 min}} d m_{23}^2 \, |\M^{(a)}_{f \bar f}|^2
\label{Gff}
\ee
where $m_{12}$ and $m_{23}$ are kinematic variables defined as
\bea
 m_{12}^2 &=& M_a^2 + m_f^2 -2 M_a E_{f} \\
 m_{23}^2 &=& M_a^2 + M_S^2 -2 M_a E_S
\eea
and the extremes of integration are
\bea
 {m_{12}^\text{2 max}} &=& \( M_a - m_f \)^2 \\
 {m_{12}^\text{2 min}} &=& \( M_S + m_f \)^2
\eea
\bea
 {m_{23}^\text{2 max}} &=& \( E_2^*+E_3^* \)^2 - \(\sqrt{E_2^{*2}-m_f^2} - \sqrt{E_3^{*2}-m_f^2} \)^2 \\
 {m_{23}^\text{2 min}} &=& \( E_2^*+E_3^* \)^2 - \(\sqrt{E_2^{*2}-m_f^2} + \sqrt{E_3^{*2}-m_f^2} \)^2
\eea
where
\bea
 E_2^* &=& \frac{m_{12}^2 - M_S^2 + m_f^2}{2 m_{12}} \\
 E_3^* &=& \frac{M_a^2 - m_{12}^2 - m_f^2}{2 m_{12}}
\eea
and $m_f$ is the fermion mass. The Feynman amplitude of the process is
\footnote{In the $R_\xi$ gauge, $\xi=1$.}
\bea
 &&\!\!\!\!\!\!\!\!\!\!\!\!
   \M_{f \bar f}^{(a)} =- i e C_\g^{(a)} k^\m \bar u_S \g_5 \[ \g_\m,\g_\n \] u_\l \frac{\eta^{\n \r}}{k^2} \bar u_f \g_\r v_f + \nn\\
 &&\phantom{\!\!\!\!\!\!\!\!\!\!\!\!\M_{f \bar f}^{(a)} =}
   - i \frac{g_{Z_0}}{2} C_{Z_0}^{(a)} k^\m u_S \g_5 \[ \g_\m,\g_\n \] u_\l \frac{\eta^{\n \r}}{k^2-\MZO^2}
                  \bar u_f \g_\r (v_f^{Z_0} - a_f^{Z_0} \g_5) v_f          \label{Mff}\\
 &&\phantom{\!\!\!\!\!\!\!\!\!\!\!\!\M_{f \bar f}^{(a)}}
   = - i k^\m \bar u_S \g_5 \[ \g_\m,\g_\n \] u_\l \[e q_f C_\g^{(a)} \frac{\eta^{\n \r}}{k^2} \bar u_f \g_\r v_f +
                    \frac{g_{Z_0}}{2} C_{Z_0}^{(a)} \frac{\eta^{\n \r}}{k^2-\MZO^2} \bar u_f \g_\r (v_f^{Z_0} - a_f^{Z_0} \g_5) v_f \] \nn
\eea
where $q_f$, $v_f^{Z_0}$ and $a_f^{Z_0}$ are respectively the electric charge, vectorial and axial coupling of the
$Z_0$, $C_\g^{(a)}$ and $C_{Z_0}^{(a)}$ are given respectively in eq. (\ref{Cg}) and (\ref{CZ0}).
More details can be found in Appendix~\ref{appnn}.

\section{Gaugino Radiative Decay. Branching Ratio}
\begin{figure}[t]
\centering
\includegraphics[scale=0.35]{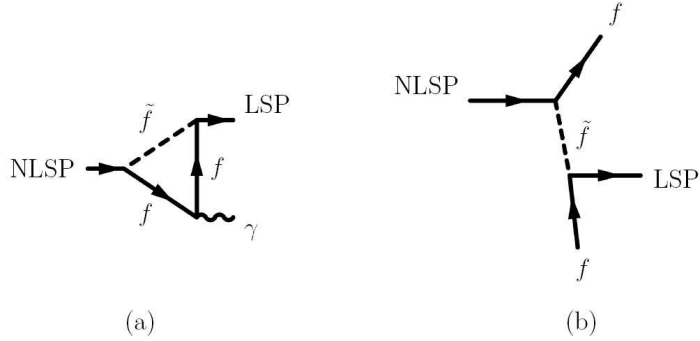}
\caption{Some diagrams contributing to NLSP decay in the MSSM: (a) NLSP $\to$ LSP $\g$ (b) NLSP $\to$ LSP $f \bar f$ }
\label{MSSMdecays}
\end{figure}

\begin{figure}[p]
 \centering
 \includegraphics[scale=0.42]{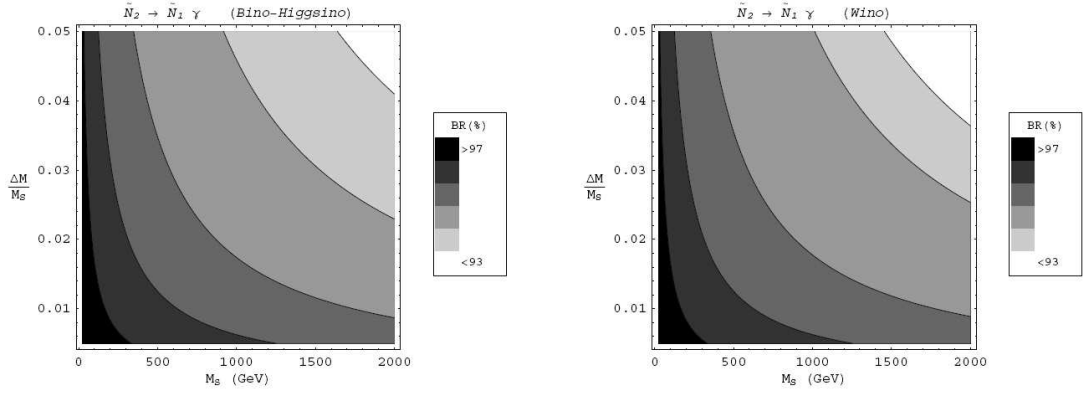}
\caption{Branching Ratio for the gaugino radiative decay. On the x-axis we have the axino mass, while on the y-axis we have the percentage mass gap.}
\label{BRplot2}
\end{figure}

\begin{figure}[p]
 \centering
 \includegraphics[scale=0.42]{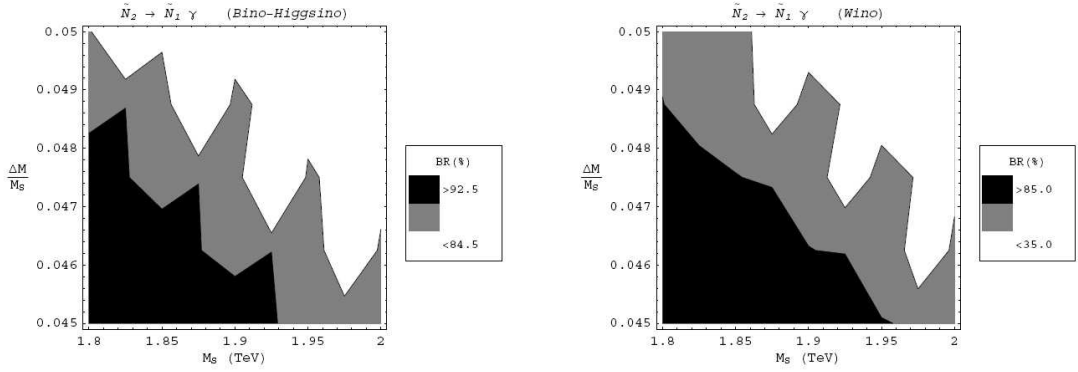}
\caption{Branching Ratio for the gaugino radiative decay. Zoom in the high $\D M$ region.}
\label{BRplotzoom2}
\end{figure}

\begin{figure}[p]
 \centering
 \includegraphics[scale=0.42]{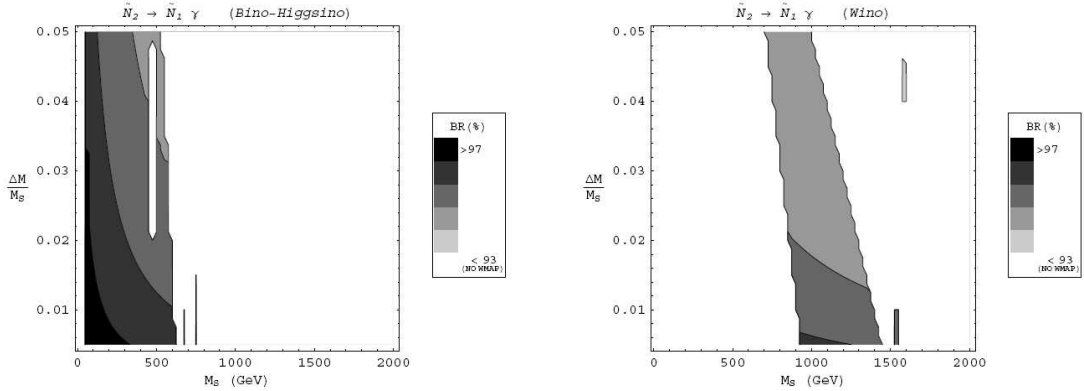}
\caption{Branching Ratio for the gaugino radiative decay. WMAP allowed region.}
\label{BRWMAP2}
\end{figure}

In this section we summarize the results about the BR of the gaugino radiative decay.
We consider an axino LSP and the NLSP to be either a bino-higgsino mixture or a nearly pure wino.
We solve numerically the integral (\ref{Gff}) and we express the result as a function of the axino mass $M_S$ and the mass gap $\Delta M/M_S$. The axino mass ranges from $25$ GeV up to $2$ TeV while $\Delta M/M_S$ ranges from $0.5\%$ up to $5\%$.
The results are shown in the contour plots in Fig.~\ref{BRplot2}-\ref{BRWMAP2}
as functions of $M_S$ and $\D M/M_S$.
There is no dependence on $\MZp$ and on the anomaly involved in the process
$\cA^{(a)}$, since they factorize in the decay rates in $b_2^{(a)}$ (see eq. (\ref{bsds})) and they cancel out in the BR computation.
It is worth noting that there is no dependence on the gaugino fraction of the neutralino.
As expected the BR is very high both for the bino-higgsino and wino case and the corresponding plots (Fig.~\ref{BRplot2})
have no substantial differences.
In a wide region of the parameters the BR is higher than $93\%$ since the contribution coming from Fig.~\ref{binodecay}b is
kinematically forbidden and the correction coming from Fig.~\ref{binodecay}c is only few percents
 (a second order process in perturbation theory).
The situation is very different from the CMSSM case where the BR is lower than 1\%, so it is never dominant \cite{Baer:2002kv}.
 In the unconstrained MSSM we expect the one loop process $\tilde N_2 \to \tilde N_1 \g$ (see fig.~\ref{MSSMdecays}(a)) to be suppressed with respect to
 tree level process $\tilde N_2 \to \tilde N_1 f \bar f$ (see fig.~\ref{MSSMdecays}(b)) although the tree level decay contribution can be lowered by a
suitable choice of the free parameters.


The high $\D M$ regions plotted in Fig.~\ref{BRplotzoom2} deserves a comment.
In this region the contribution coming from Fig.~\ref{binodecay}b now is
kinematically allowed but suppressed because $\D M$ is not very high with respect to $\MZO$.
In this case an important role is played
by the rotation factors (\ref{Rg}) and (\ref{RZ}). In the bino-higgsino NLSP case they favor the radiative decay, while in the wino NLSP case they favor the
decay with the $Z_0$ production and so there is and interplay between the kinematic suppression and the rotation factor enhancement.
The result is that for the bino-higgsino NLSP case, the radiative process is still the dominant one, while for the wino NLSP case the
decay rate for $Z_0$ production can be higher than the radiative decay rate.

The WMAP allowed regions of parameters \cite{Fucito:2008ai} are plotted in Fig.~\ref{BRWMAP2}.
The white region represents either BR$<93\%$ or the WMAP forbidden region. Only for the wino case we have a
WMAP allowed white region which is a tiny vertical strip at $M_S \sim 1.6$ TeV and
$\D M/M_S \geq 0.04$.
We see that there is a huge difference in the allowed region for the two cases. The bino-higgsino case is allowed
at low masses ($M_S < 700$ GeV) while the wino case is allowed at high masses ($M_S > 700$ GeV).
In both cases the vertical strip $M_S < 50$ GeV is excluded by the lower mass limit
on the MSSM neutralinos \cite{Amsler:2008zzb} and the $Z_0$ production region is forbidden.
\begin{figure}[t]
 \centering
 \includegraphics[scale=0.4]{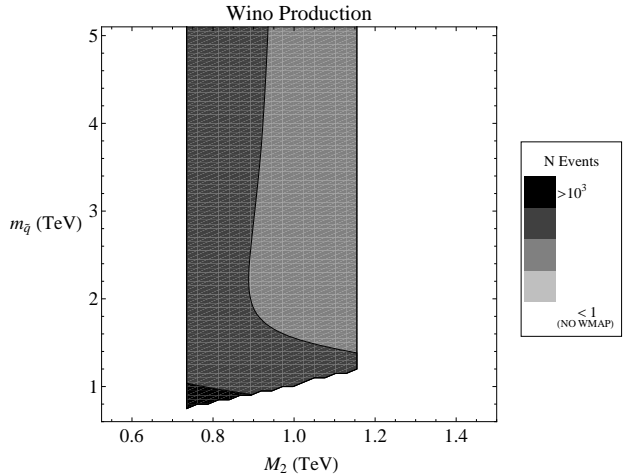}
\caption{Number of wino produced inside the WMAP allowed region.
         Black scale means more than $10^3$ then we lower the order of magnitude by a factor of 10 down to 1 (white scale).}
\label{Nwinos}
\end{figure}

By assuming a center of mass energy of 14 TeV and an integrated luminosity of $100\, {\rm fb}^{-1}$, we give an estimate of the number of NLSP produced.
We focus on the wino case. The leading production processes are \cite{Baer:2006rs}: $q \bar q \to \l_2 \l_2$, $q g \to \l_2 \tilde q$,
$q \bar q \to \l_2 \tilde g$, $q \bar q' \to \l_2 \l_2^\pm$, where $\l_2^\pm$ is the wino-like chargino. The parton cross sections were integrated using the
parton distribution function (PDF) package \cite{Alekhin:2006zm}. As an example we fixed the percentage mass gap at 5\% and the gluino mass is $M_3 \sim 3 M_2$
(typical in anomaly mediation). The wino-like chargino mass is the wino NLSP mass, since the mass degeneracy between the two states.
As a simplifying assumption, we considered an universal squark mass $m_{\tilde q}$ for all the squarks involved.
The dominant process is  $q g \to \l_2 \tilde q$ because of the gluon PDF.
The number of NLSP produced is given in Fig.~\ref{Nwinos} as a function of the wino mass $M_2$ and the squarks mass $m_{\tilde q}$.
Since the BR is almost close to one this is also the number of photons in the final state.
The number of events is always greater than 10 while in the low mass region the number can be greater than 1000.
It should be stressed that our results
are not a distinctive signature of a SUSY anomalous $U(1)'$ model but
they are rather a consequence of the absence of a direct coupling of the LSP with
(s)fermions.
However this result can be combined with direct
\cite{ourpaper} and indirect searches \cite{Armillis:2008vp} of
anomalous $Z'$ triangle interactions, measurements of the
fermionic couplings by $Z'$ decay width and forward-backward
asymmetries at LHC. We postpone a detailed analysis of these points in a forthcoming paper~\cite{toappear}.

\section{Conclusions}
We computed the BR for the radiative decay of the NLSP in the model described in \cite{ourpaper}.
Motivated by \cite{Fucito:2008ai}, we considered an axino LSP and the NLSP to be alternatively a bino-higgsino mixture or a nearly pure wino. In both cases
we found a very high BR~($>93\%$).
This result is different from the CMSSM case where the BR is typically very low~($<1\%$).
The corresponding WMAP allowed region are very different:
low axino masses~($M_S<700$ GeV) for the bino-higgsino case, high masses~($M_S>700$ GeV) for the wino case.
This result could be used to discriminate between the two options for a possible decay in LHC.
Anyway we will come back to this point in a forthcoming paper where we analyze in details the number of
events for the gaugino radiative decay inside LHC \cite{toappear}.

\vskip 2cm
\begin{flushleft}
{\large \bf Acknowledgments}
\end{flushleft}

\noindent The authors would like to thank prof. Francesco Fucito
for useful discussions and Daniel Ricci Pacifici for the help
given us in the wino production computation. A. R. would like to
thank Prof. Hermann Nicolai and the MPI for Gravitational Physics
(AEI) in Potsdam for hospitality and financial support. This work
was partially supported by the ESF JD164 contract.

\vskip 2cm
\appendix

\section{$\l^{(a)} \to \psi_S f \bar f$ decay rate} \label{appnn}
The amplitude in given by eq. (\ref{Mff}). The corresponding square modulus is
\bea
 |\M^{(a)}_{f \bar f}|^2 &=& -32 \[
 T_a \(\frac{ a_f C_{Z_0}^{(a)} g_{Z_0} }{k^2-\MZO^2} \)^2
 +T_v \( \frac{ 2 C_\g^{(a)} e q_f}{k^2 }+ \frac{C_{Z_0}^{(a)} g_{Z_0} v_f}{k^2-\MZO^2} \)^2 \]
\label{nunuampl}
\eea
with $q_f$ the electric charge of the fermion, $e$ the corresponding coupling constant,
$v_f(a_f)$ the vectorial (axial) coupling of the fermion to $Z_0$, $g_{Z_0}$ the corresponding coupling constant,
$m_f$ the fermion mass, $k$ the virtual gauge boson momentum, $C_{\g/Z_0}^{(a)}$ given by (\ref{Cg}), (\ref{CZ0}) and
\bea
 T_v &=&m_f^4 (p_a p_S)
     -M_1 M_S \Big[2 m_f^4+3 (p_f p_{\bar f}) m_f^2+(p_f p_{\bar f})^2\Big]+\nn\\
&&-(p_f p_{\bar f})
   \Big[(p_a p_f) (p_f p_S)+(p_a p_{\bar f}) (p_{\bar f} p_S)\Big] +m_f^2 \Big[(p_a p_S) (p_f p_{\bar f})+\nn\\
&&-2 (p_a p_f) (p_f p_S)-(p_a p_{\bar f})
   (p_f p_S)-(p_a p_f) (p_{\bar f} p_S)-2 (p_a p_{\bar f}) (p_{\bar f} p_S)\Big] \nn\\
 T_a &=& \Big[(p_a p_{\bar f}) (p_f p_S)+(p_a p_f) (p_{\bar f} p_S)\Big] m_f^2+M_1 M_S
   \[m_f^4-(p_f p_{\bar f})^2\]+\nn\\
&&-(p_f p_{\bar f}) \Big[(p_a p_f) (p_f p_S)+(p_a p_{\bar f})(p_{\bar f} p_S) \Big]
\eea
where
$p_a$, $p_{S}$, $p_f$ and $p_{\bar f}$ are the gaugino, axino and
SM fermions 4-momenta respectively. The factor $32$ contains also the $1/2$ factor coming from the average of the
gaugino spin states. The scalar products of the momenta are given by
\bea
 p_a p_S        &=& M_a E_S \nn\\
 p_a p_f        &=& M_a E_f \nn\\
 p_a p_{\bar f} &=& M_a (-E_f - E_S + M_a) \nn\\
 p_f p_S         &=& M_a \(E_f + E_S\) - \frac{M_a^2+M_S^2}{2} \nn\\
 p_f p_{\bar f}  &=& \frac{1}{2} \( M_a^2 + M_S^2 -2 E_S M_a - 2 m_f^2 \) \nn\\
 p_{\bar f} p_S  &=& \frac{1}{2} \( M_a^2 - M_S^2 -2 E_f M_a \) \nn\\
 k^2             &=& M_a^2 + M_S^2 -2 E_S M_a
\eea
and the integration variables are introduced by the equations
\bea
 E_f &=& \frac{M_a^2 + m_f^2 - m_{12}^2}{2 M_a} \nn\\
 E_S &=& \frac{M_a^2 + M_S^2 - m_{23}^2}{2 M_a}
\eea

\end{document}